\documentclass[journal]{IEEEtran}

\ifCLASSINFOpdf
\else
   \usepackage[dvips]{graphicx}
\fi
\usepackage{url}

\usepackage{graphicx}
\usepackage{amsfonts}
\usepackage{amsmath}

\begin{document}

\title{Using maximum weighted likelihood to derive Lehmer and Hölder mean families}

\author{Djemel Ziou 
\thanks{D\'epartement d'informatique, Universit\'e de Sherbrooke,  Sherbrooke, Qc., Canada J1K 2R, Djemel.Ziou@usherbrooke.ca }}

\maketitle

\begin{abstract}
In this paper, we establish the links between the  Lehmer and H\"older  mean families and  maximum weighted likelihood estimator.  Considering the regular one-parameter exponential family of probability density functions, we show that the  maximum weighted likelihood of the parameter is a generalized weighted mean family from which  Lehmer and H\"older mean families are inferred.    Some of the outcomes obtained provide a probabilistic interpretation of these mean families and could therefore broaden their uses in various applications.
\end{abstract}


\IEEEpeerreviewmaketitle

\section{Introduction}
Central tendency refers to a mechanism for choosing an element that best resembles the elements of a set. 
The best-known central tendencies  for centuries are the arithmetic, geometric and harmonic means, grouped under the name of Pythagoreans.  Recently, it has been shown that each of these implements a data selection rule~\cite{Ziou22a}.  Lehmer and H\"older means families, which generalize the Pythagoreans, are other useful central tendencies~\cite{deCarvalho16,Sluciak15,Burrows86,Borwein98,Beckenbach50,Stolarsky96}.   Each family is made up of an infinity of means and each is specified by a value of a real parameter. For example, arithmetic, geometric, and harmonic means can be derived from either the  Lehmer or H\"older mean families by setting specific values for the real parameter. Lehmer family is used in differential evolution~\cite{Das17}, neural networks~\cite{Terziyan22},  extreme events estimation~\cite{Penalva20}, and depressive disorders characterization~\cite{Ataei22}. H\"older family is used in several  areas, including  electricity, information retrieval, finance, health, filtering, neural network, and human development assessment~\cite{Tripathi11,deCarvalho16}.  In what follows the terms "mean", "mean family", and central tendency are  indifferently used.
Some of these means are maximum likelihood estimators (MLEs). Indeed, there is a one-to-one correspondence between a probability density function (PDF), a mean and a MLE of the location parameter or the scale parameter. For example, normal is the only location-based PDF whose arithmetic mean is the MLE of its location parameter and exponential PDF is the one whose arithmetic mean is the MLE of its scale parameter. The reader can find more about MLE characterization in~\cite{Duerinckx14}. However, one-to-one correspondences between  Lehmer and H\"older means and MLEs do not yet appear to have been established. 
Thanks to the notion of weighted likelihood (WL), in this paper we will establish correspondences between these means, regular one-parameter exponential family of PDFs, and maximum weighted likelihood estimators (MWLEs)~\cite{Ziou23a}. Precisely, we will show that: 1) these  means  are MWLE for a subclass of this family of PDFs; 2) the MWLE does not only depend on a PDF as shown in existing MLE characterization studies, but it also depends on the relevance of the data. To our knowledge, the derivation of Lehmer and H\"older means as MWLE   have not yet been done before. The correspondences provide a probabilistic interpretation of these means and would therefore make it possible to broaden their use in the many fields where the MLE is used. The paper is organized as follows. The next section outlines the two central tendencies. In section~\ref{Sect3}, we derive the MWLEs and show how they are linked to these central tendencies. Section~\ref{Sect4} presents case studies.

\section{Lehmer and H\"older Central tendencies}

\begin{figure}
	\begin{center}
		\includegraphics[height=4cm, width=4cm]{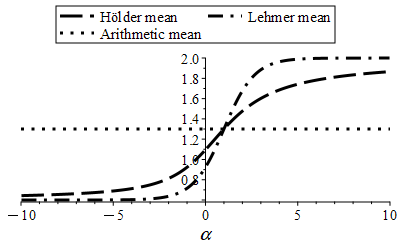}{a)}
		\includegraphics[height=4cm, width=4cm]{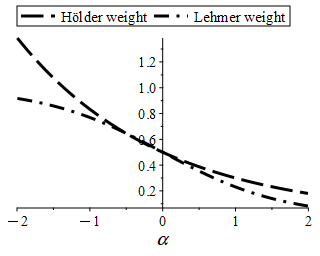}{b)}
		\includegraphics[height=4cm, width=4cm]{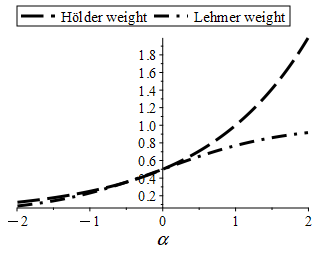}{c)}
		\caption{(a) Three means between $0.6$ and $2$ as function of $\alpha$. The arithmetic mean is displayed as a basis for comparison. (b) The weights  $v_h(0.6)$ and $v_l(0.6)$  as a function of $\alpha$.  (c) The weights  $v_h(2)$ and $v_l(2)$ as a function of $\alpha$.   
		}
		\label{MeansPlot}
	\end{center}
\end{figure}

Let us consider the  IID data ${\cal X}=x_1, \cdots, x_n$, where $x_i \in  \mathbb{R}_{\ge 0}$. Several formulas exist for calculating the mean of ${\cal X}$, and most of them are  a particular case of the generalized f-mean, known also as the Kolmogorov mean~\cite{Burrows86}:
\begin{equation}
	K_f=f^{-1}(\frac{1}{n}\sum_i f(x_i))
	\label{Kolmogorov}
\end{equation}
where the function $f : \mathbb{R}_{\ge 0} \rightarrow \mathbb{R}_{\ge 0}$  is continuous and increasing. The  H\"older  family of means $H_\alpha$ is  particular cases of the generalized f-mean, where $\alpha \in \mathbb{R}$. It is obtained by setting  $f(x)=x^\alpha$: 
\begin{equation}
	H_\alpha=(\frac{1}{n} \sum_i  x_i^\alpha)^{1/\alpha}
\end{equation}
Due to its connection with the $p$-norm, it is widely used in information technology, finance, health,  and human development assessment, pattern recognition, among many other areas~\cite{Tripathi11,deCarvalho16,Oh16}.  Lehmer $L_\alpha$ is an alternative to the f-mean and it is  given by: 
\begin{equation}
	L_\alpha=\frac{\sum_i x_i^\alpha}{\sum_j   x_j^{\alpha-1}}
\end{equation}
It is used in  differential evolution~\cite{Das17}, neural networks~\cite{Terziyan22},  extreme events estimation~\cite{Penalva20}, and depressive disorders characterization~\cite{Ataei22}.
Both families are bounded by the smallest and greatest values of ${\cal X}$ and  are  continuously non-decreasing  functions with respect to $\alpha$. 
For the purpose of comparison  between both, Fig.~\ref{MeansPlot}.a represents  the Lehmer and H\"older  central tendencies of the two equiprobably
numbers $x_1=0.6$ and $x=2$ as function of $\alpha$ and the arithmetic mean as a basis. The Lehmer  is greater than the  H\"older  when $\alpha > 1$,  lower for $\alpha < 1$, and equal when $\alpha=-\infty, 1, +\infty$.  The Pythagorean central tendencies are particular cases;  the geometrical mean $H_0 = L_{0.5}$ when $n=2$,  the arithmetic mean $H_1=L_1$, and the harmonic mean 
$H_{-1}=L_0$.  Moreover, the slope of the Lehmer  is higher than the H\"older, i.e. the Lehmer  reaches  the lowest and highest values more quickly. Both are smaller than the arithmetic mean when $\alpha<1$,
greater when $\alpha > 1$ and equal when $\alpha=1$. The reader can find more details about these means in~\cite{Beckenbach50,Burrows86}. 

An important issue we want to address is the explanation of the data selection embedded  in the central tendencies.   
Drawing inspiration from the interpretation of the mean described in~\cite{Ziou22a}, we provide two data selection mechanisms  embedded  in each of  Lehmer and H\"older means. The first is the w-weight $w(x)$ encoding   knowledge about the observations ${\cal X}$, such as the frequency  or a prior knowledge. The higher $w(x_i)$, the more $x_i$ contributes to the central tendency. The H\"older and Lehmer w-weighted means are written, respectively, as:  
\begin{equation}
	H_\alpha=(\frac{\sum_i w(x_i) x_i^\alpha}{\sum_j w(x_j)})^{1/\alpha}~~~\mbox{and}~~~L_\alpha=\frac{\sum_i w(x_i) x_i^\alpha}{\sum_j w(x_j) x_j^{\alpha-1}}
	\label{Arithmean0}	
\end{equation}
The second mechanism is the v-weight $v(x)$ based on the value $x$; that is to say that a measurement contributes to the calculation of the central tendency according to its value $x$. More precisely, the H\"older and Lehmer v-weighted means can be written, respectively, as:  
\begin{equation}
	H_\alpha=(\sum_i v_h(x_i) x_i)^{1/\alpha}~~~\mbox{and}~~~L_\alpha=\sum_i v_l(x_i) x_i
	\label{Arithmean}
\end{equation}

where  $v_l(x_i)=w(x_i) x_i^{\alpha-1}/\sum_j w(x_j) x_j^{\alpha-1}$ and $v_h(x_i)=w(x_i) x_i^{\alpha-1}/\sum_j w(x_j)$. Note that while the sum of the v-weights is equal to one (i.e. $ \sum_i v_l(x_i)=1$) in the case of Lehmer, it is not the same for H\"older because $\sum_i v_h(x_i)=H^{\alpha-1}_{\alpha-1}$ for $\alpha \ne 1$. The link between the two means is  straightforward  $L_\alpha=H_\alpha^\alpha/H^{\alpha-1}_{\alpha-1}$.  To better illustrate these v-weights, let us consider again the two equiprobably numbers $x_1=0.6$ and $x_2=2$.   Fig.~\ref{MeansPlot}.b depicts  $v_h(0.6)=0.6^\alpha/2$ and $v_l(0.6)=0.6^\alpha/ ( 0.6^\alpha+2^\alpha)$    and Fig.~\ref{MeansPlot}.c  $v_h(2)= 2^\alpha/2$ and $v_l(2)= 2^\alpha/ ( 0.6^\alpha+2^\alpha) $ as function of $\alpha$. The function $v_z(0.6)$ is decreasing and $v_z(2)$ is increasing in $\alpha$, where $z$ is either $l$ or  $h$. In other words, the relevance of $x_1=0.6$  (resp. $x_2=2$)   is decreasing (resp. increasing) when $\alpha$ is increasing. Hence, one of the data selection mechanisms embedded in the two means involves increasing values above one and weakening values below one.

\section{Maximum weighted likelihood estimates}
\label{Sect3}
Since the introduction of the  version we know today by R. A. Fisher over a century ago, the maximum likelihood has been widely studied and much has been written~\cite{Duerinckx14,Hald99,Burrows86,Miura11,Myung03,Rossi18}. Among the outcomes, we can state that the MLE  has asymptotically the lowest variance among all unbiased estimators. For the one-parameter natural exponential family, it  was shown that there is a one-to-one correspondence between a PDF and the MLE of the location or the scale parameter~\cite{Duerinckx14}. For example, the normal  is the only location-based PDF whose arithmetic mean is the MLE of the location parameter, and the exponential PDF is the one whose  arithmetic mean  is the MLE of its scale parameter.  It should be noted that these  outcomes and many others were obtained when measurements  have the same weight. Let us go further considering the weighted data ${\cal D}= {\cal X} \times {\cal U} = (x_1,u(x_1)), \cdots, (x_n,u(x_n))$ where  $x_i \in \mathbb{R}_{\ge 0}$ and $u(x_i) \in \mathbb{R}_{> 0}$ is a weight of the observation $x_i$.  The relationship between the weight function $u(x)$ and the previously introduced w-weight and v-weight will be further elaborated.
Let us consider that  the data ${\cal X}$,  is sampled from  a one-parameter  exponential family  of a continuous variable $x \in \mathbb {R}_{\ge 0}$ whose  PDF is given by:
\begin{equation}
	f(x|\theta)=a(x) exp(\eta(\theta) T(x)-H(\theta)) 
	\label{pdfgexp}
\end{equation}
where   $H(\theta)=ln \int_{\mathbb {R}_{\ge 0}} a(x) exp(\eta(\theta) T(x)) dx$ is the normalizer, $a(x)$ the basis measure  associating non-negative values to $x$ regardless of $\theta$,  $T(x)$ is referred to as  a sufficient statistic, and $\eta(\theta)$  the parametrization function which we consider  increasing and continuous  in $\theta$ (just multiply both $\eta(\theta)$ and $T(x)$ by minus one, if $\eta(\theta)$ is decreasing).  For there to be a definite order relation on the data, $T(x)$ should be continuous and  monotonic.  By setting  each of $a(x)$, $T(x)$, and $\eta(\theta)$ to  a specific value, several existing PDFs can be derived from Eq.~\ref{pdfgexp}  such as the normal, beta, gamma, and Weibull. Note that we limit ourselves to identifiable models.

 We will use  WL to establish the relationship between Lehmer and Holder's central tendencies with the MWLE. The WL was proposed by Feifang Hu and which consists of integrating observation weights into the Fisher likelihood~\cite{Hu94}. Under the IID assumption, for the PDFs in~ Eq. \ref{pdfgexp}, the WL of the weighed data ${\cal D}$ is given by:
\begin{equation}
	L(\theta)= \prod_i (a(x_i) exp(\eta(\theta) T(x_i) - H(\theta)))^{u(x_i)} 
	\label{LH}
\end{equation}
The positive weight function $u(x)$  must be $\theta$ free. Equating the first derivative with respect to $\theta$ of the log-WL to zero and resolving, leads us to write:
\begin{equation}
	r(\theta)=\sum_i u(x_i) T(x_i)/\sum_i u(x_i)
	\label{CP}
\end{equation}
where $r(\theta)=H'(\theta)/\eta'(\theta)$.  Since the derivative of $H(\theta)$ wrt $\theta$ is $H'(\theta)=\eta'(\theta)  E_\theta(T(x))$, then   $r(\theta) =E_\theta(T(x))$, where $E_\theta(T(x))$ is the expectation of $T(x)$. 
By using  the Lebesgue dominated convergence theorem, the  derivative of $r(\theta)$ wrt $\theta$ is given by:
\begin{equation}
	\begin{split}
	r'(\theta)= &\frac{d E_\theta(T(x))}{d \theta}
	=\frac{d }{d \theta} \int_{\mathbb {R}_{\ge 0}} T(x) a(x) exp(\eta(\theta) T(x)- 
\\ 	& H(\theta)) dx = \eta'(\theta) V_\theta(T(x))
	\end{split}
\end{equation}
The functions $r'(\theta)$   and  $\eta'(\theta)$ have the same sign because the variance $V_\theta(T(x))$ is positive. Since $\eta(\theta)$ is assumed  increasing and continuous, then the function $r(\theta)$  is also increasing. It follows that  $r(\theta)$ is invertible, and, therefore,  the critical point is:
\begin{equation}
	\theta = r^{-1}(\sum_i u(x_i) T(x_i)/\sum_i u(x_i))
	\label{CriticalP}
\end{equation}
This formula, less specialized than the generalized f-mean in Eq.~\ref{Kolmogorov}, will  be used later to derive the Lehmer and H\"older means. Now, let's determine the nature of the critical point. The second derivative of the log-WL  is:
\begin{equation}
	L''(\theta)=\eta''(\theta) \sum_i u(x_i) T(x_i) - H''(\theta) \sum_i u(x_i) 
	\label{2L}
\end{equation}
Straightforward computation from $H(\theta)$ leads to  $H''(\theta)=\eta'(\theta) V_\theta(T(x))+\eta''(\theta) E_\theta(T(x))$. At the critical point, the Eq.~\ref{2L} is reduced to $L''(\theta)=-\eta'(\theta) V_\theta(T(x)) \sum_i u(x_i)$. Since $\eta(\theta)$ is increasing function, then $L''(\theta)$ is negative and, therefore the critical point in Eq.~\ref{CriticalP} is a maximum. 

Another important issue concerns the selection of data, namely the function $u(x)$ to use.  In our case, the objective is to use the functions $u(x)$ which make it possible to derive   Lehmer and H\"older means as  MWLE.  For this, comparing   Eq.~\ref{CP} and  Lehmer mean in Eq.~\ref{Arithmean0} gives a solution (not necessarily unique) $u(x)=w(x) x^{\alpha-1}$ and $T(x)=r(\theta) x/|r(\theta)|$; that is  the MWLE  in Eq.~\ref{CP} is a function of Lehmer mean. A subclass of PDFs from the one-parameter exponential family leading to a function of Lehmer mean as MWLE has the form $a(x) exp(\eta(\theta) r( \theta) x/|r(\theta)|-H(\theta))$ when $u(x)= w(x) x^{\alpha-1}$.
Comparing Eq.~\ref{CP} and H\"older mean in Eq.~\ref{Arithmean0}  raised to $\alpha$ power givesa solution (not necessarily unique) $u(x)=w(x)$ and $T(x)=r(\theta) x^\alpha/|r(\theta)|$ ; that is
the MWLE  in  Eq.~\ref{CP} is a function of H\"older mean. A subclass of PDFs from the one-parameter exponential family leading to a function of H\"older mean as MWLE has the form $a(x) exp(\eta(\theta) r( \theta) x^\alpha/|r(\theta)|-H(\theta))$ when $u(x)= w(x)$. Note that, these comparisons show that $u(x)$ is nothing other than the w-weight in Eq.~\ref{Arithmean0}. 

\section{Case studies}
\label{Sect4}
To illustrate, in table~\ref{Laws} we provide seven PDFs, where four have two parameters. For each of these four PDFs, assuming that the shape parameter $\alpha$ is known, we derive the MWLE of the scale parameter $\theta$. 
In this table, $u(x)=x^\beta$ and the MWLE is a function of the Lehmer central tendency in the cases of exponential,  gamma, and inverse gamma. If $\beta=0$ (i.e. $u(x)=1$), the MWLE, which reduces to the MLE, is a function of the H\"older central tendency in the cases of Weibull, generalized half-normal, and half-normal. 
It should be noted that in the examples of table~\ref{Laws}, there is a one-to-one correspondence between the MWLE of the location parameter or the scale parameter and the pair (PDF, $u(x)$). This observation is a generalization of the established result indicating that for a one-parameter natural exponential family, there is a one-to-one correspondence between a PDF and the MLE of the location parameter or the scale parameter. In summary, WL made it possible to produce a probabilistic interpretation of the central tendencies of Lehmer and H\"older.  The links between the Lehmer central tendency and the MWLE as well as between  the H\"older central tendency and the MWLE  do not appear to have been established before.

\begin{figure}
	\centering 
	\includegraphics[height=4cm, width=5.5cm]{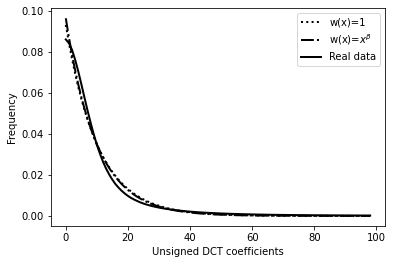}{a)} 
	\hspace{1cm}
	\includegraphics[height=4cm, width=5.5cm]{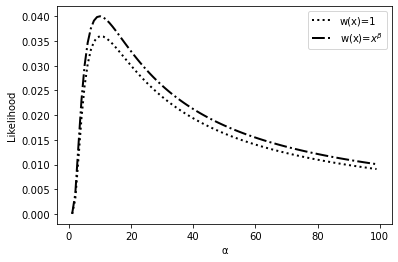}{b)}
	\caption{(a) Real and estimated histograms  by using the Weibull PDF and the weight functions $u(x)=x^\beta$ and $u(x)=1$. The estimated $\beta=-0.02$ and $\alpha=1$. (b) The corresponding WL. }
	\label{HFE}
\end{figure}

\begin{figure}
	\centering 
	\includegraphics[height=4cm, width=5.5cm]{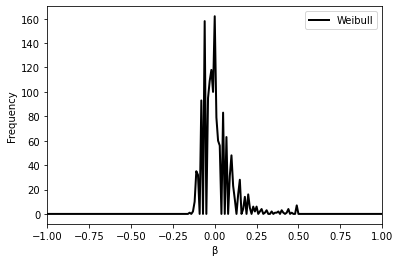}  
	\caption{The histograms of the parameter $\beta$. }
	\label{Beta}
\end{figure}

We will now estimate the MWLE and parameter $\beta$ of the weight function $u(x)$ for histogram fitting.  Histogram fitting is widely used in statistical inference and machine learning. For this, histograms constructed from the real data are fitted by the Weibull PDF using MWLE and the weighting function $u(x)=x^\beta$. We consider two distinct cases that are ($\beta = 0$, $\alpha \ne 0$) and ($\beta \ne 0$, $\alpha = 1$), where $\alpha$ is the shape parameter of Weibull PDF (See Table~\ref{Laws}). The former leads to the H\"older central tendency as the MLE of Weibull PDF, while  the later leads to Lehmer central tendency  as MWLE of exponential PDF. Note that the Weibull PDF with $\alpha=1$ is the exponential PDF.  The shape parameter $\alpha$ of Weibull and the parameter $\beta$ are estimated by minimizing the mean square error (MSE) between the histogram and the Weibull PDF.
Specifically, MWLE and MSE are executed alternately until convergence. Data are 1500 histograms of the absolute value of the coefficients of the discrete cosine transform (DCT) of 1500 gray-level images of indoor and outdoor scenes, taken day and night, extracted from the Unsplash collection\footnote{https://unsplash.com/collections}. The DCT coefficients are  widely used in image compression algorithms and in image quality assessment~\cite{Kerouh17,Manfredi11}.  Fig.~\ref{HFE} presents an example of a DCT histogram  fitted by the Weibull PDF with  the estimated  $\alpha=1$ (i.e. exponential PDF). The  estimated $\beta=-0.02$. The estimated MSE for 55.6\% of the histograms is smaller when employing the Lehmer central tendency (i.e., $\beta \ne 0$, $\alpha=1$) compared to 44.4\% of the histograms when using the H\"older central tendency (i.e., $\beta=0$, $\alpha \ne 0$).
Figure ~\ref{Beta} depicts the histogram of $\beta$ values, showcasing a positively skewed distribution. Zero values correspond to the H\"older mean, while non-zero values suggest that employing the Lehmer mean results in a more precise fit for the DCT histograms. To conclude, the use of Lehmer's or Holder's means within the framework of the MWLE provide a valuable explanation. These means provide statistical tools to evaluate and calculate estimates in scenarios where MWLE is used, thereby improving our understanding and allowing us to draw meaningful conclusions from the data. This connection makes it possible to identify the subclasses of PDFs leading to these means like MWLE. Moreover, it facilitates a more complete interpretation of the results obtained through MWLEs, strengthening their applicability and relevance in various fields of study. 

\begin{table}
	\begin{tabular}{|l|l|l|l|l|}
		\hline
		PDF &  $\eta(\theta)$ & $H(\theta)$ & $T(x)$  & MWLE (Eq.~\ref{CP})\\
		\hline
		Exponential &  $\theta$ & $-ln(\theta)$ & $-x$   & $\frac{1}{\theta}=\frac{\sum_i x_i^{\beta+1}}{\sum_i x_i^{\beta}}$\\  \hline   
		Weibull  &  $\theta^\alpha$ & $-\alpha ln(\theta)$ &$-x^{\alpha}$   & $\frac{1}{\theta^\alpha}=\frac{\sum_i x_i^{\alpha+\beta}}{\sum_i x_i^\beta}$\\ \hline  
		St. L-normal  &   $\frac{1}{2} \theta^2$ & $-ln(\theta)$  &$-ln^2 x$  & $\frac{1}{\theta^2}=\frac{\sum_i x_i^\beta ln^2(x_i)}{\sum_i x_i^\beta }$   \\  \hline 
		H-normal  &  $\frac{1}{2} \theta^2$ & $-ln(\theta)$   & $-x^2$ &     $\frac{1}{\theta^2}=\frac{\sum_i x_i^{\beta+2}}{\sum_i x_i^{\beta}}$   \\  \hline
		Gen. H-normal &  $\frac{1}{2} \theta^{2\alpha}$ & $-\alpha ln(\theta)$ &$-x^{2\alpha}$ &   $\frac{1}{\theta^{2\alpha}}=\frac{\sum_i x_i^{2\alpha+\beta}}{ \sum_i x_i^{\beta}}$   \\  \hline     
		Gamma  &  $\theta$ & $-\alpha ln(\theta)$ & $-x$ &     $\frac{\alpha}{\theta}=\frac{\sum_i x_i^{\beta+1}}{\sum_i x_i^{\beta}}$   \\   \hline
		Inv. Gamma  &  $\theta$ & $- \alpha ln(\theta)$&$-\frac{1}{x}$  & $\frac{\alpha}{\theta}=\frac{\sum_i x_i^{\beta-1}}{ \sum_i x_i^{\beta}}$   \\  \hline
	\end{tabular}
	\caption{Example of PDFs within the exponential family along with their associated MWLEs of the parameter $\theta$ are presented. The shape parameter $\alpha$  is assumed to be known. The same weight function $u(x)=x^\beta$ is used for all the PDFs. Abbreviation: Standard (St), Inverse (Inv), Generalized (Gen), Half (H). }
	\label{Laws}
\end{table}

\bibliographystyle{elsarticle-num}

\end{document}